\documentclass[twoside,draft]{article}
\usepackage{latexsym,e-journal}

\usepackage[letterpaper]{geometry}
\usepackage[utf8]{inputenc}
\usepackage[T1]{fontenc}
\usepackage[english]{babel}

\usepackage{amssymb,amsfonts,amsmath}

\usepackage[perpage,symbol*]{footmisc}
\usepackage[final]{graphicx}
\usepackage{pstricks}
\usepackage{cite}

\usepackage[varg]{txfonts}

\begin{document}

\renewcommand{\refname}{References}
\renewcommand{\tablename}{\small Table}
\renewcommand{\figurename}{\small Fig.}
\renewcommand{\contentsname}{Contents}

\def \pteptitle {Searching for the Feynman diagram IIc}
\def \ptepauthor {Oliver Consa}

\twocolumn[%

\begin{center}
\renewcommand{\baselinestretch}{0.93}
{\Large\bfseries \pteptitle
}\par
\renewcommand{\baselinestretch}{1.0}
\bigskip
Oliver Consa\\ 
{\footnotesize  Department of Physics and Nuclear Engineering, Universitat Politècnica de Catalunya \\ 
Campus Nord, C. Jordi Girona, 1-3, 08034 Barcelona, Spain\rule{0pt}{8pt}\\
E-mail: oliver.consa@gmail.com
}\par
\medskip
{\small\parbox{11cm}{%
The calculation of the electron g-factor was carried out in 1950 by Karplus and Kroll. Seven years later, Petermann detected and corrected a serious error in the calculation of a Feynman diagram. Although it's hard to believe, neither the original calculation nor the subsequent correction was ever published. Therefore, the entire prestige of QED and the Standard Model depends on the calculation of a single Feynman diagram (IIc) that has never been published and cannot be independently verified. In this article we begin the search for any published recalculation of this Feynman diagram IIc that allows us to independently validate the theoretical calculation.
}}
\smallskip
\end{center}] {%

\setcounter{section}{0}
\setcounter{equation}{0}
\setcounter{figure}{0}
\setcounter{table}{0}
\setcounter{page}{1}

\markboth{\ptepauthor. \pteptitle}{\ptepauthor. \pteptitle}

\markright{\ptepauthor. \pteptitle}
\section{The big problem}
\markright{\ptepauthor. \pteptitle}

\subsection{Renormalization}
 
The Standard Model of Particle Physics brings together two different physical theories: Electroweak Theory (EWT) and Quantum Chromodynamics (QCD). For decades a "Grand \linebreak Unification Theory (GUT)" has been unsuccessfully sought to integrate both theories into one unified theory. 

Both QCD and EWT are mainly mathematical theories. The aim is to identify a set of gauge symmetries for each theory that allows a concrete mathematical formulation to be obtained. EWT forms a SU(2) x U(1) symmetry gauge group while QCD forms a SU(3) symmetry gauge group. The theory is considered correct if the theoretical values obtained with these mathematical formulas coincide with the experimental values obtained with particle colliders.

Both QCD and EWT are based on and completely dependent on the validity of quantum electrodynamics (QED), developed by Feynman, Schwinger, and Dyson. QED in turn is a quantum field theory (QFT). QFT emerged in the 1930s in an attempt to quantify the electromagnetic field itself. But QFT has a serious problem. All calculations give the same result: Infinity. 

In the 1940s, QED developers managed to solve the infinities problem using a technique called "Renormalization.". Many methods can be used to eliminate these infinities, but the main ones are:

\begin{itemize}
\item Substitution: replacing a divergent series with a specific finite value that has been arbitrarily chosen (for example, the energy of an electron).
\item Separation: separating an infinite series into two components, one that diverges to infinity and another that converges to a finite value. Eventually, the infinite component is ignored and only the finite part remains.
\item Cut-off: focusing on an arbitrary term in the evolution of a series that diverges to infinity and ignoring the rest of the terms of the series.
\end{itemize}

\begin{figure}[ht]
	\centering
	\includegraphics[scale=0.6]{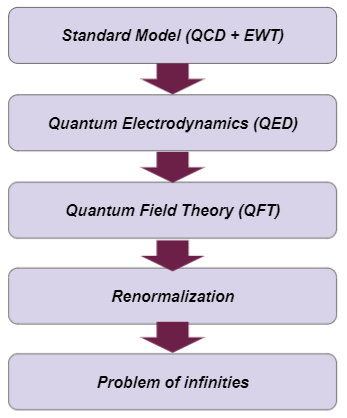}
	\vspace*{-3mm}
\caption{Layers of logical dependencies.}
\end{figure}

As an example of the use of these Renormalization techniques we can look at the calculation of the Casimir effect \cite{Casimir}. The equation of the Casimir effect depends on the Riemann function. However, the Riemann function is defined only for positive values, since for negative values the Riemann function diverges to infinity. 

\begin{equation} 
\frac{F_c}{A} = \frac{d}{da}\frac{<E>}{A} =- \frac{\hbar c \pi ^2}{2 a^4} \ \zeta(-3) = \frac{\hbar c \pi ^2}{20 a^4} \ \zeta(-1)
\end{equation}

In particular the Riemann function of -1 corresponds to the value of the sum of all positive integers. Applying a Renormalization technique, the Indian mathematician \linebreak Ramanujan came to the conclusion that the sum of all positive integers is not infinity but -1/12 \cite{Integers}. And this is precisely the value that is used in the equation of the Casimir effect.

\begin{equation} 
\zeta(-1) = \sum _{n=1}^{\infty }{n} = 1+2+3+4+5+ ... =  \frac{-1}{12}
\end{equation}

\begin{equation} 
\frac{F_c}{A} = \frac{\hbar c \pi ^2}{20 a^4} \left (\frac{-1}{12} \right ) = - \frac{\hbar c \pi ^2}{240 a^4}  
\end{equation} 

Despite being one of the main creators of QED, Feynman was not very convinced about Renormalization:\textit{“The shell game that we play is technically called ’renormalization’. But no matter how clever the word, it is still what I would call a dippy process! Having to resort to such hocus-pocus has prevented us from proving that the theory of quantum electrodynamics is mathematically self-consistent. It’s surprising that the theory still hasn’t been proved self-consistent one way or the other by now; I suspect that renormalization is not mathematically legitimate.”} \cite{Feynmanbook}

For his part, Dirac was always clearly against these techniques: \textit{“I must say that I am very dissatisfied with the situation because this so-called ’good theory’ does involve neglecting infinities which appear in its equations, ignoring \linebreak them in an arbitrary way. This is just not sensible mathematics. Sensible mathematics involves disregarding a quantity when it is small – not neglecting it just because it is infinitely great and you do not want it!.”} \cite{Dirac}

Today the scientific community accepts these renormalization techniques as fully legitimate. But if Dirac was right and renormalization is not a legitimate mathematical technique, then the Standard Model, EWT, QCD, QED and all theories based on QFT would be incorrect and worthless. 

\subsection{QED precision}

The entire credibility of the renormalization techniques is \linebreak based on its level of precision of the theoretical value with respect to the experimental value. As an example, the electron g-factor offers an  impressive level of precision of 12 decimal places:

\begin{itemize}
\item Experimental value \cite{Gabrielse}: $1.001,159,652,180,73(28)$
\item Theoretical value \cite{Kinoshita}: $1.001,159,652,182,032(720)$ 
\end{itemize}

In 1970 Brodsky and Drell summarised the situation in their paper "The present status of the Quantum Electrodynamics" as follows: \textit{“The renormalization constants are infinite so that each calculation of a physical quantity has an infinity buried in it. Whether this infinity is a disease of the mathematical techniques of perturbation expansions, or \linebreak whether it is symptomatic of the ills accompanying the idealization of a continuum theory, we don't know. Perhaps there is a "fundamental length" at small distances that regularizes these divergences (...). Quantum electrodynamics has never been more successful in its confrontation with experiment than it is now. There is really no outstanding discrepancy despite our pursuing the limits of the theory to higher accuracy and smaller (...) however, and despite its phenomenal success, the fundamental problems of renormalization in local field theory and the nature of the exact solutions of quantum electrodynamics are still to be resolved.”} \cite{Drell}

It seems inconceivable that using an incorrect theory we can obtain the correct results with an unprecedented level of precision. And it is extremely unlikely that this finite theoretical value coincides with the experimental value by pure chance. Therefore, the only reasonable explanation is that renormalization techniques must be mathematically \linebreak legitimate even though we cannot prove it at the moment.

\begin{figure}[ht]
	\centering
	\includegraphics[scale=0.6]{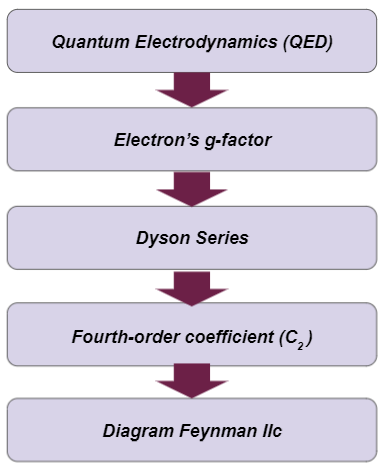}
	\vspace*{-3mm}
\caption{Layers of logical dependencies.}
\end{figure}

\subsection{Dyson series}

Quantum Electrodynamics (QED) is considered the most accurate theory in the history of science. However, this impressive precision is limited to a single experimental value: the anomalous magnetic moment of the electron (g-factor). 

According to Quantum Electrodynamics (QED), the theoretical value of the electron g-factor is obtained by calculating the coefficients of a number series called the Dyson series [4]. Each coefficient in the series requires the calculation of an increasing number of Feynman diagrams.  

\begin{equation}
g = C_1 \left(\frac{\alpha}{\pi}\right) + 
C_2 \left(\frac{\alpha}{\pi}\right)^2 + 
C_3 \left(\frac{\alpha}{\pi}\right)^3 + 
C_4 \left(\frac{\alpha}{\pi}\right)^4 +
C_5 \left(\frac{\alpha}{\pi}\right)^5 ...
\end{equation}

The first coefficient in the Dyson series is the Schwinger factor and has an exact value of 0.5. The second coefficient was initially calculated in 1950 by Karplus and Kroll \cite{KK},and it was corrected in 1957 by Petermann \cite{Peter}, who obtained a result of -0.328. The rest of the coefficients in the Dyson series were calculated many decades later with the help of supercomputers. 
\begin{equation} 
g =  1 + \frac {1}{2} \left(\frac {\alpha}{ \pi}\right) - 0,328 \left(\frac {\alpha}{ \pi}\right)^2 = 1,0011596 
\end{equation}

This result of the $C_2$ coefficient of the Dyson series \linebreak (fourth-order coefficient) was decisive for the acceptance of the renormalization techniques proposed by Feynman, \linebreak Schwinger, and Tomonaga, who received the Nobel Prize in 1965 for the development of QED. It can therefore be considered the most relevant theoretical calculation in modern physics.

\subsection{Feynman Diagram IIc}

The error in the calculation of $C_2$ discovered by Petermann was found in the calculation of the Feynman diagram IIc. 

\begin{figure}[ht]
	\centering
	\includegraphics[scale=0.6]{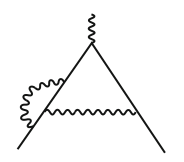}
	\vspace*{-3mm}
\caption{Feynman diagram IIc.}
\end{figure}

According to the Karplus and Kroll’s, original calculation, the value of diagram IIc was -3.178 while in the Petermann correction the value of diagram IIc was -0.564.
\linebreak

[Karplus \& Kroll]
\begin{equation}
  II_c = -\frac{323}{24} + \frac{31}{9} \pi ^2 -  \frac{49}{6} \pi ^2 ln (2) +  \frac{107}{4} \zeta(3) = -3.178
\end{equation}

[Petermann]
\begin{equation}
II_c = -\frac{67}{24} +  \frac{1}{18} \pi ^2 +  \frac{1}{3} \pi ^2 ln (2) -  \frac{1}{2} \zeta(3) = -0.564
\end{equation}

However, hard to believe, neither the original calculation carried out in 1950 by Karplus and Kroll nor the subsequent correction of Petermann was ever published. Therefore, the entire legitimacy of the Standard Model and the QED depends on the calculation of a single Feynman diagram (IIc) that has never been published and cannot be independently verified. \cite{Consa2}

\section{Searching for the missing calculation}

\subsection{Barbieri \& Remiddi}

At this point, we set out on a mission to find the missing calculus of the Feynman diagram IIc. We assume that given the seriousness of the situation, someone must have recalculated previously this Feynman diagram and published it years ago.

After a long search we believe we found the paper we were looking for. It is a paper published in 1972 and written by Remiddi among other authors \cite{Remiddi}. 
Remiddi is one of the most prestigious researchers in the calculation of the electron g-factor because in 1996 he published the definitive analytical value of the $C_3$ coefficient (sixth-order coefficient). 

The paper is a long 93-page document entitled "Electron form factors up to fourth order". It was published in 1972 by Barbieri and Remiddi. According to the authors: \textit{“This paper is devoted to the analytic evaluation of the two form factors of the electromagnetic vertex of the electron in quantum electrodynamics, up to fourth order of the perturbative expansion (...) [Calculation] of the fourth-order form factors can also be found in the literature. They are the famous fourth-order anomalous magnetic moment evaluated by Petermann and Sommerfield (...). Such values are obviously reproduced in this paper. (...) Calculations are done in the framework of the usual Feynman-graph expansion of the S-matrix in the interaction representation, using the Feynman gauge for the photon propagator. The relevant graphs for second-order and fourth-order radiative corrections are shown (..). The approach is dispersive, and the discontinuities of the various Feynman graphs are obtained by means of the Cutkosky rules”.} \cite{Remiddi}

From this introduction we understand that Barbieri and Remiddi performed a recalculation of the Feynman diagrams corresponding to the Fourth-order coefficient ($C_2$) and they confirmed the results obtained by Petermann. 

\begin{figure}[ht]
	\centering
	\includegraphics[scale=0.6]{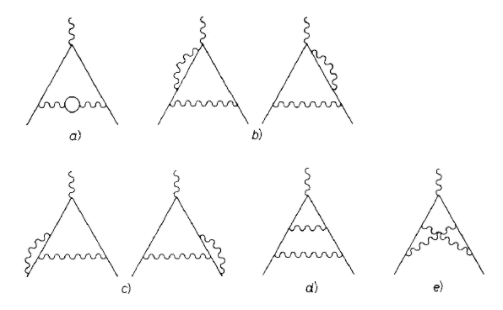}
	\vspace*{-3mm}
\caption{$C_2$ Feynman diagrams.}
\end{figure}

The authors identify the Feynman IIc diagram as the "c" diagram in Figure 4, divide it into two symmetrical diagrams and give it the descriptive name of "Corner Graphs". The result shown in the paper is identical to that published by Petermann in 1957.

In the 93 pages of the paper, the authors describe several of the techniques they have used to renormalize the divergences that appear in the calculations and how they have overcome the problems they have encountered. On the specific calculations, the authors state the following: \textit{“Once these problems are mastered, a very long and complicated algebra is also needed to do in practice the calculation. Fortunately, the major part of it, like traces, straightforward algebraic manipulations, book-keeping of analytic formulae, integrations by parts, differentiations and so on, was done by computer, using the program SCHOONSCHIP of VELTMAN, without whose continuous and determinant help the present work could hardly have been accomplished.”} \cite{Remiddi}

That is, they used a computer program to perform the mathematical calculations, but they did not publish the code of the programs used, so, again, it is not possible to replicate the calculations.

Considering the date of the paper (1972) it is quite plausible to assume that there are no more calculations, since it was considered unnecessary to carry out more checks of the $C_2$ coefficient. Fortunately, in the paper itself the authors identify two other independent calculations of the  $C_2$ coefficient. One published in 1960 by Smrz and Uleha and the other published in 1962 by Terentiev.  

\subsection{Smrz \& Uleha}

We obtain the paper published in 1960 by these two Czech researchers \cite{Smrz}. It is a short paper of 2 pages where the situation generated in 1957 by Petermann's correction is explained. The paper indicates that the difference between the original Feynman IIc diagram calculation of Karplus and \linebreak Kroll with respect to the one performed by Petermann is excessive. The authors states that they performed an independent calculation of the Feynman IIc diagram and obtained exactly the same result as Petermann. 

\textit{“Since the considerable difference between the original value of the magnetic moment (Karplus \& Kroll \cite{KK}) and the values calculated later (Petermann \cite{Peter}) originates in the calculation of the contribution from the third diagram, only the value of this contribution was determined by the standard technique and the above regularization in the infra-red region. The contribution from the third diagram (-0.564) is in complete agreement with Petermann's value.”} \cite{Smrz}

Unfortunately, when searching for the reference of the work we note that it has not been published either: \textit{"Smrz P.: Diploma thesis, Faculty of Tech. and Nucl. Physics, Prague 1960, unpublished."} 

Just another unpublished paper claiming to have calculated the Feynman IIc diagram but with no one to review it. 

\subsection{Terentiev}

We obtained a copy of the paper published by Terentiev in 1962. The paper contains about 50 pages \cite{Terentiev}. The paper is only in Russian and there is no English translation. We identify the equation "60" of the paper as the $C_2$ coefficient of the Dyson series, with the same expression and value obtained by Petermann and Sommerfield.

\begin{equation} 
g =  1 + \frac {1}{2} \left(\frac {\alpha}{ \pi}\right) - 0,328 \left(\frac {\alpha}{ \pi}\right)^2 = 1,0011596 
\end{equation}

Analyzing the document, we interpret that this equation is the result of the sum of nine other equations identified as equations 22, 24, 27, 31, 33, 47, 51, 58 and 59. There are nine equations instead of the five Feynman diagrams of Karplus and Kroll and none of the these equations correspond to the Feynman Diagram IIc. 

However, it is not necessary to carry out a more in-depth analysis of the paper. On the first page of the \linebreak Barbieri-Remiddi paper we can read a reference to Terentiev's paper: \textit{“Actually, dispersion relations are used in the Terentiev work only to write down suitable multiple integral representations, which are in general manipulated to get the final result, without explicitly evaluating the discontinuities. The problem of infra-red divergences has been further overlooked, and many of the intermediate results are wrong, even if somewhat ad hoc compensations make the final result correct.”}

\renewcommand{\arraystretch}{1.3}
\begin{table}[ht]
\begin{tabular}{|c|c|p{4.3cm}|}
\hline
Year & Author &  Status \\
\hline
1950 & Karplus \& Kroll & Wrong and Unpublished \\
\hline
1957 & Petermann & Right but Unpublished \\  
\hline
1957 & Sommerfield & Right but using Green Functions instead of \linebreak Feynman Diagrams \\
\hline
1960 & Smrz \& Uleha & Right but Unpublished \\
\hline
1962 & Terentiev & Wrong intermediate results with ad hoc compensations to make the final result correct \\
\hline
1972 & Remiddi & Right but Unpublished \linebreak Computer calculation  \\
\hline
\end{tabular}
\caption{\label{tab:Calculations} fourth-order coefficient calculation.}
\end{table}

\section{Summary}
Incredible as it may seem to believe, the most important calculation in the history of modern physics was published in 1950 by Karplus and Kroll and turned out to be completely incorrect. The error was not detected until 7 years later by Petermann and Sommerfield. Neither the original calculation nor the subsequent correction was ever published. Therefore, the entire legitimacy of the Standard Model and the QED depends on the calculation of a single Feynman diagram (IIc) that has never been published and cannot be independently verified.

In this paper we have detected three other published recalculations of the fourth-order coefficient of the g-factor. The detailed calculations of two of them were also not published (Barbieri-Remiddi and Smrz-Uleha). In the third calculation performed by Terentiev, serious errors were detected 10 years after the original publication. Erroneous intermediate results manipulated with ad hoc compensations to obtain the correct final result. 

Our search has been extensive, so we believe that there are no other published calculations of the Feynman IIc diagram. The only line of investigation that remains open would be to find the source code of the computer programs that are currently used to carry out this type of calculation.

\begin{flushright}\footnotesize
1 September 2021
\end{flushright}

\vspace*{-6pt}
\centerline{\rule{72pt}{0.4pt}}
}

\end{document}